# Posture Recognition in the Critical Care Settings using Wearable Devices

Anis Davoudi, PhD[1], Patrick J. Tighe, MD[2], Azra Bihorac, MD[3], Parisa Rashidi, PhD[1]
[1]Department of Biomedical Engineering, University of Florida, Gainesville, FL, USA;
[2]Department of Anesthesiology, University of Florida, Gainesville, FL, USA;
[3]Department of Nephrology, University of Florida, Gainesville, FL, USA

**Abstract**

*Low physical activity levels in the intensive care units (ICU) patients have been linked to adverse clinical outcomes. Therefore, there is a need for continuous and objective measurement of physical activity in the ICU to quantify the association between physical activity and patient outcomes. This measurement would also help clinicians evaluate the efficacy of proposed rehabilitation and physical therapy regimens in improving physical activity. In this study, we examined the feasibility of posture recognition in an ICU population using data from wearable sensors.*

**Introduction**

Patients in the intensive care unit (ICU) spend a significant amount of time lying in bed (1, 2). Previous research has shown the negative impact of decreased physical activity and lack of movement on patients' recovery and hospital outcomes (3). Low physical activity during the ICU stay is associated with adverse outcomes such as a higher risk of mortality, delirium, more prolonged ICU and hospital stay, less favorable discharge location, and worse post-discharge functional status and cognitive status (4, 5).

Early rehabilitation and physical therapy have been proposed and implemented to increase patients' physical activity in the ICU. These interventions have been linked to improved outcomes among ICU patients (3, 6, 7). An accurate, continuous, and granular approach to assess patients' physical activity can additionally validate and provide a comparison of the efficacy of different physical therapy methods.

Wearable accelerometer devices have been used extensively for activity recognition and posture detection in different populations (8-10). Currently available wearable accelerometer devices, such as Fitbit, are lightweight and unobtrusive. They can accurately measure the acceleration of human movement. These devices have been used to study different aspects of physical activity in sports and clinical research in daily life (11-13). Wearable accelerometer devices have previously been used in clinical studies in hospital wards and ICUs (1, 14-16). In hospital settings, they have been used to monitor the well-being of health care providers, for example, nurses' physical activity and fatigue (17). Wearable accelerometer devices also have been used to study patients' fall risk and mobility, recovery, mental health, and pain level (14, 18-20). For example, previously wearable devices have been used to show the difference in physical activity levels in different patient populations, detect delirium and subtypes, detect sepsis, and measure sleep quality during hospital stays.

The accelerometer data analysis methods can be primarily classified into conventional methods and deep learning methods. Conventional methods rely on manually extracted features (aka feature engineering), while deep learning approaches extract features automatically from data. Deep learning models learn the patterns in the data by optimizing network parameters with iterative fine-tuning of the weights in the model layers. On the other hand, conventional machine learning models, such as random forest and support vector machines (SVM), rely on engineered features. In essence, the model 'learns' the relationship between the features and the activity class during the training step. Then, based on the learned relationship, it will determine the class label of unseen test data points. Building useful features for the activity recognition task requires domain knowledge and examination of previous literature to understand what features will be helpful in each specific physical activity research task.

Calculation of features usually consists of first extracting a window (e.g., a segment of the timeseries consisting of two seconds of data), and then calculating the mentioned in each window. The duration of the time window used in calculation of the features is important and its effect on the performance of the models has been investigated in previous studies in non-hospital settings. However, its effect has not been investigated in the ICU/hospital settings where the patients move slowly, which affects their frequency.

Another important consideration when using wearable accelerometer devices is where the device is worn on the body. Different device positions on the body are optimal for different research tasks. While many step counters rely on hip and ankle positions, the wrist has been a favorite position in many research applications and daily use devices (e.g., Fitbit) because the wearable device feels like a watch and is not perceived as inconvenient by the wearer.

In fact, smartwatches, such as those manufactured by Apple and Samsung, are actual watches that also have embedded accelerometers that can report on physical activity levels. As these watches can collect patient-reported data via other applications related to the user's status, they have also been used in clinical research (21, 22).

In this study, we evaluated the use of data from wearable accelerometer devices to detect patients' physical activity levels in the ICU. The objectives of this study were to evaluate the performance of classifiers using wearable accelerometer devices for posture recognition in the ICU, and to determine the effect of device position on the performance of these classifiers. To the best of our knowledge, this paper presents the first study that used wearable accelerometer devices for posture recognition in the ICU.

**Methods**

This study was carried out in the surgical ICU of the University of Florida Health Hospital in Gainesville, Florida. It was approved by the University of Florida Institutional Review Board (IRB 201400546), and all methods were performed following the relevant guidelines and regulations. Written, informed consent was obtained from patients or their surrogates before enrollment in the study.

Participants:

We recruited patients >18 years old who were admitted to the ICU with an expected stay of more than 48 hours without intubation. For each patient, accelerometer data were collected for up to one week, or until discharge or transfer from the ICU, whichever occurred first. Cohort inclusion and exclusion criteria are described in another study (1). Patient demographic information and relevant EHR data were obtained from the University of Florida Health Hospital data repository.

Data Collection:

- Video system:

For patients who consented to video recording in the room, a system with a commercial camera with an attached computer system to control the recording was placed in front of the wall facing the patient's bed. Video data collection was performed for up to seven days or until discharge from the ICU, as per the one which happened earlier. Nurses and other caregivers were able to pause or stop the recording or delete the recorded videos according to patient request. The recorded video files were transferred to a secure server after the data collection for each patient was completed. A trained annotator randomly sampled four 15-minute video clips per day and annotated each frame to report each sample's postures.

- Wearable Accelerometers:

Patients who consented to wearing wearable accelerometer devices were asked to wear up to three accelerometer devices: on the right ankle, on the wrist, and on the upper arm. Similar to the video data collection, accelerometer data collection was continued for up to seven days or discharge from the ICU. Nurses and other caregivers could remove the devices in case of discomfort, during bathing, or if needed for clinical care routines. Recorded raw data for each patient were downloaded and converted using the ActiLife toolbox and saved in a secure server. Only data from devices worn on the wrist and ankle were used in the analyses.

- Analysis:
  - Accelerometer Features

Timeseries recordings (at 100 Hz) were converted to 10 Hz data since human movements typically do not exceed 10 Hz (23). Several time and frequency domain features discussed in the literature were extracted from the timeseries recordings for devices worn on the wrist and ankle, using window durations of two seconds for feature extraction (Table 2).

  - Classification:

Classification of posture labels was performed using random forest models (24), a classification algorithm using many decision trees in which the outcome of the classification model is determined based on the majority vote of the decision trees. Random forest models have shown strong performance in previous research (8, 25) for posture recognition using features extracted from wearable accelerometer devices. Three separate classes of models were used: 1) models using accelerometer data from the wrist position, 2) models using accelerometer data from the ankle position, and 3) models using accelerometer data from the wrist and ankle positions combined. We used nested cross-validation (8 by 3) to separate the training, validation, and testing datasets based on participant. For robust analysis, we used leave-one-out cross-validation, resulting in eight outerfolds such that, in each fold, data from one patient was used to test the model. A validation dataset was used to tune model parameters, and the final models were tested on the unseen test data in each fold. Final reported performance metrics were based on the performance of the models on the test data. In addition to the original models, to address the expected imbalance in dataset, we used downsampling, and Synthetic Minority Over-sampling TEchnique (SMOTE) (26) to tune model parameters and train the models. In downsampling, the majority class is undersampled to balance the data classes. In SMOTE, additional samples are generated for the minority class (oversampling) to increase their number while majority class is undersampled. Increased size of the minority class in SMOTE leads to losing fewer samples from the majority class.

**Table 1.** Description of features extracted from accelerometer data.

| Variable | Definition |
|---|---|
| MVM | Mean of VM in the window |
| SDVM | Standard deviation of VM in the window |
| Mean angle of acceleration relative to vertical on the device (mangle) | Sample mean of the angle between x axis and VM in the window |
| SD of the angle of acceleration relative to vertical on the device (sdangle) | Sample standard deviation of the angles in the window |
| Covariance | Covariance of the VM in the window |
| Skewness | Skewness of the VM in the window |
| Kurtosis | Kurtosis of the VM in the window |
| Entropy | Entropy of the VM in the window |
| Coefficient of variation (CV) | Standard deviation of VM in the window divided by the mean, multiplied by 100 |
| Corr (x, y) | Correlation between X axis and Y axis |
| Corr (y, z) | Correlation between Y axis and Z axis |
| Corr (x, z) | Correlation between X axis and Z axis |
| Percentage of the power of the vm that is in 0.6-2.5 Hz (p625) | Sum of moduli corresponding to frequency in this range divided by sum of moduli of all frequencies |
| Dominant frequency of vm (df) | Frequency corresponding to the largest modulus |
| Fraction of power in vm at dominant frequency (fpdf) | Modulus of the dominant frequency/sum of moduli at each frequency |

**Results**

Dataset:

We included data collected from patients who wore both ankle and wrist-worn devices and had both sitting and lying activity labels during their windows of labeled data. We only included sitting and lying activity labels because the standing level was only present for one of the patients during the labeling windows. We excluded data samples in which 1) the patient was outside the camera view (because the label of physical activity label could not be determined), 2) either wrist or ankle accelerometer data were missing, or 3) non-wear time was detected. Non-wear time was determined as continuous zero movements for longer than two hours during the daytime. We also only included data recorded during the daytime shift (7.00 a.m.-7.00 p.m.) to have clear and similar visibility across all patient videos for annotation.

Our final dataset included eight patients, and included 127,688 segments, each two-seconds long. Table 2 shows brief demographics of the cohort that appears in the analysis. There were 120,554 segments in the class lying and 7,134 segments in the class sitting. Figures 1-2 show the distribution of the features between the two classes of lying and sitting for ankle and wrist device positions, respectively. All variables except for V9 were significantly different between the two activity classes for data from the ankle-worn devices, while all variables except V18 had significantly different distributions between the two classes when using data from the wrist-worn devices ($p<0.01$).

Posture Recognition Analysis:

Table 3 shows the results of the posture recognition models for the three classes of the data using 2-seconds window segments for feature extraction. Tables 4-5 show the improvement in the performance of the models when using downsampling and SMOTE to deal with the significant imbalance in the data labels.

**Table 2.** Cohort description.

| Variable | Distribution |
|---|---|
| Age, mean (sd) | 62.3 (17.5) |
| Sex: female, N (%) | 5 (62.5) |
| BMI, mean (sd) | 24.6 (5.5) |

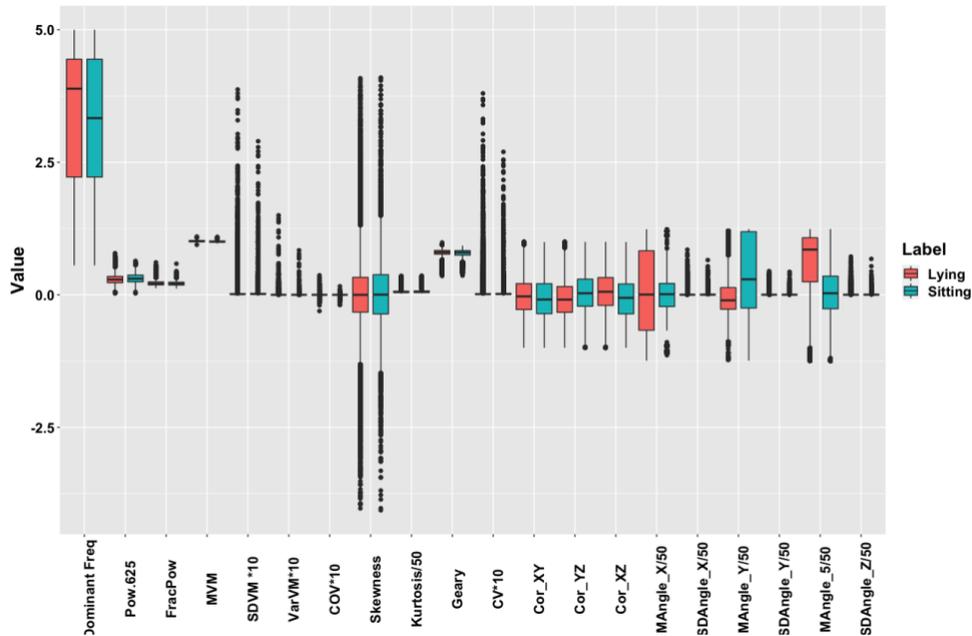

**Figure 1.** Distribution of extracted features from ankle-worn accelerometer devices for lying and sitting postures. The range of variables was adjusted for better visibility in the same figure.

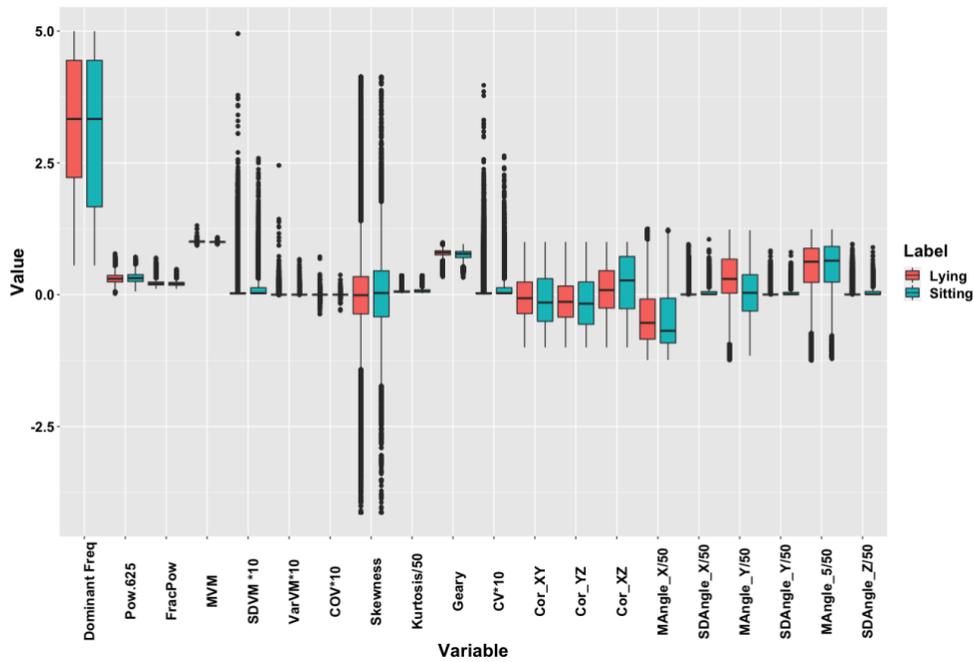

**Figure 2.** Distribution of extracted features from wrist-worn accelerometer devices for lying and sitting postures. The range of variables was adjusted for better visibility in the same figure.

**Table 3.** Performance of models in the classification of sitting and lying postures in ICU patients.

| Model | Accuracy | F1-score | Precision | Recall | Balanced Accuracy |
| --- | --- | --- | --- | --- | --- |
| Wrist positions | 0.81 | 0.02 | 0.01 | 0.03 | 0.45 |
| Ankle positions | 0.94 | 0.33 | 0.40 | 0.28 | 0.63 |
| Wrist and ankle positions | 0.94 | 0.35 | 0.44 | 0.29 | 0.64 |

**Table 4.** Performance of models in the classification of sitting and lying postures in ICU patients, using downsampling approach to address data imbalance.

| Model | Accuracy | F1-score | Precision | Recall | Balanced Accuracy |
| --- | --- | --- | --- | --- | --- |
| Wrist positions | 0.54 | 0.09 | 0.05 | 0.42 | 0.49 |
| Ankle positions | 0.93 | 0.43 | 0.37 | 0.50 | 0.73 |
| Wrist and ankle positions | 0.91 | 0.36 | 0.30 | 0.46 | 0.70 |

**Table 5.** Performance of models in the classification of sitting and lying postures in ICU patients, using SMOTE approach to address data imbalance.

| Model | Accuracy | F1-score | Precision | Recall | Balanced Accuracy |
|---|---|---|---|---|---|
| Wrist positions | 0.62 | 0.07 | 0.04 | 0.25 | 0.45 |
| Ankle positions | 0.93 | 0.43 | 0.39 | 0.47 | 0.71 |
| Wrist and ankle positions | 0.93 | 0.35 | 0.35 | 0.35 | 0.66 |

**Discussion**

Summary of Findings:

We used data collected from wearable accelerometer devices worn on the ankle and wrist to study the feasibility of posture recognition in the ICU. Data samples were labeled as sitting or lying using video data recorded in the patient's room during the data collection period. Random forest classifiers were trained using the features extracted from the accelerometer devices to classify the posture as sitting or lying. Classifier models trained on the wrist were not successful at classifying patient postures (with balanced accuracy of 0.45 for original models, and balanced accuracies of 0.49 and 0.45 for models using downsampling and SMOTE, respectively). Classifier models trained on data from the ankle-worn device were more successful at classifying patient postures (with balanced accuracy of 0.63 for original models, and balanced accuracies of 0.73 and 0.71 for models using downsampling and SMOTE, respectively). Multi-source classifiers incorporating data from both ankle and wrist positions performed similarly to classifiers using data from ankle-worn devices, with a balanced accuracy of 0.64 for original models and 0.70 and 0.66 for models using downsampling and SMOTE, respectively. The improvement in the models' performance in all metrics seen when using downsampling and SMOTE, shows the importance of having more samples in the minority class (sitting). While our data shows the nature of physical activity levels in real ICU settings in which the patients spend the majority of their time in bed, larger samples of the minority class will help improve model performance.

It should be noted that movements that occur in lying and sitting postures while in bed are not very different in ICU patients, as in both cases the legs remain in a straight position. Rather, only the angle of the back changes. As a result, the data from the ankle-worn devices may not be sufficient to differentiate well between sitting in bed and lying in bed. Moreover, sitting on a chair in the ICU is also not very different from lying, as would be the case for sitting versus lying in normal settings. The patients' legs are usually drawn up, leading to smaller differences in the angles between upper and lower legs, as is generally the case for normal sitting and closer to the position of the legs when lying. Wrist movement also depends on the patients' level of consciousness and may be affected both in terms of gross movements in interacting with the environment and in terms of more subtle movements and agitations. This means that the data collected from wrist-worn devices may be reporting other activities that are not relevant to posture recognition. These points show the limitations of individual sensors for posture recognition in an ICU population as opposed to their greater success in community-dwelling populations (8, 27).

Limitations and Future Directions:

One of the main limitations of the current study is the small number of participants included in the analysis. This issue mainly arises from the challenges of data collection in the ICU in terms of patient consenting to the video recording/accelerometer and research team time, as well as the crowded settings of ICU rooms, which decreases the usefulness of some of the recorded data. The huge time demand of the annotation task is a challenge in many prospective studies and leads to several other limitations, including the need for sampling data for annotation (losing the capability for analyzing the data for all the day), and less possibility of validating the annotations by having multiple annotators' labels for each sample. While the current study shows the feasibility of posture recognition in an ICU setting, future studies should examine the generalizability of the results among ICU patients with various diagnoses.

Another limitation of the study was that to provide clear and similar visibility across the patients and time points, we only included frames recorded during the daytime. However, physical activity levels of ICU patients during the nighttime will also be informative to determine their sleep/rest quality. Future studies should address this question using cameras that are equipped with night vision capabilities.

Our data only included sitting and lying physical activity levels, as we did not have a sufficiently large sample of other activity levels. Future studies should also examine the performance of the models in detecting the standing posture. One potential approach would be to use non-patient data to train the classifiers. However, because of the mentioned differences between the sitting and lying in the ICU setting as opposed to sitting and lying outside the hospital, the trained models may not necessarily transfer successfully to this population. Moreover, in our data we included both sitting in bed and sitting on a chair as 'sitting' to enhance our data size. In future work, we plan to collect a larger dataset to include sitting on a chair and sitting in bed as separate classes, as they might demonstrate different functional status for the patient and need to be quantified separately. This separate label may also improve the performance of the model by providing cleaner labels, as sitting on a chair and sitting in bed may be different in terms of the nature of movement. Another limitation of this study was that we did not examine the effect of the window length on posture recognition performance. Previous research has shown that the performance may vary based on feature extraction window duration. However, in our study we used two seconds-windows for data extraction to increase the data sample size. Future research should also examine the effect of window duration on performance.

Another promising approach for posture recognition is using computer vision, which has been researched in both non-hospital settings and hospital settings, including several studies in the ICU (1, 28-30). Incorporating more complete information from video data, such as the RGB images, as the input to the model could also improve the performance of the model in terms of better distinguishing between sitting and lying in bed. However, many issues, including patient privacy and model robustness, need to be addressed to validate this approach as a reliable method for future patient monitoring in the intelligent ICU protocol.

**Conclusion**

Posture recognition using wearable accelerometer devices has been validated in the controlled laboratory settings and mostly among healthy/community-dwelling participants. Large-scale data collection and evaluation of posture recognition in the ICU population using wearable accelerometer devices would permit enhanced examination of the possibility of reliable posture recognition.

**Acknowledgement**

PR was supported by National Science Foundation CAREER award 1750192, 1R01EB029699 and 1R21EB027344 from the National Institute of Biomedical Imaging and Bioengineering (NIH/NIBIB), R01GM-110240 from the National Institute of General Medical Science (NIH/NIGMS), 1R01NS120924 from the National Institute of Neurological Disorders and Stroke (NIH/NINDS), and by R01 DK121730 from the National Institute of Diabetes and Digestive and Kidney Diseases (NIH/NIDDK). AB was supported R01 GM110240 from the National Institute of General Medical Sciences (NIH/NIGMS), R01 EB029699 and R21 EB027344 from the National Institute of Biomedical Imaging and Bioengineering (NIH/NIBIB), R01 NS120924 from the National Institute of Neurological Disorders and Stroke (NIH/NINDS), and by R01 DK121730 from the National Institute of Diabetes and Digestive and Kidney Diseases (NIH/NIDDK). PJT was supported by R01GM114290 from the NIGMS and R01AG121647 from the National Institute on Aging (NIA). The content is solely the responsibility of the authors and does not necessarily represent the official views of the National Institutes of Health.